# Frequency Response Characteristic (FRC) Curve and Fast Frequency Response Assessment in High Renewable Power Systems

Shutang You

*Abstract*— This letter introduces a frequency response characteristic (FRC) curve and its application in high renewable power systems. In addition, the letter presents a method for fast frequency response assessment and frequency nadir prediction without performing dynamic simulations using detailed models. The proposed FRC curve and fast frequency response assessment method are useful for operators to understand frequency response performance of high renewable systems in real time.

## I. Introduction

Renewable generation is increasing in many power grids [1-6], influencing both operation and planning. Since the increase of renewable penetration reduces system inertia and governor response, power system frequency response has become a major concern of high renewable power systems. The common approach to evaluate system steady-state frequency response is a constant value $\beta$ whose unit is MW/0.1Hz. This value is often used for frequency response monitoring in wide-area measurement systems (which have many other situational awareness applications). It is also referred to as Frequency Response Obligation (FRO). This value gives a power system's real power generation increase per 0.1 Hz system frequency decrease. However, this constant value cannot depict the non-linearity caused by governor deadbands and generator headroom limits, as well as various emerging frequency-responsive resources [7-9]. Consequently, operators can hardly evaluate the system frequency response capability over a range of frequency deviations.

This letter introduces a power system frequency response characteristic (FRC) curve, as a more comprehensive metric for evaluating the frequency response capability accurately and procuring frequency response sources cost-effectively. In addition, a simplified frequency response model is proposed for fast prediction of frequency nadir, which supports the decision of under-frequency remedy strategies.

## II. Frequency Response Characteristic Curve and Fast Frequency Response Assessment

### A. Frequency Response Characteristic (FRC) Curve

The proposed FRC curve is defined as the system steady-state frequency response capability at different frequency deviation levels. It is an extension of the commonly used $\beta$ value, which can only represent a linear relation between system response and frequency deviation. Incorporating non-linear characteristics from both governor response and load damping, the FRC curve can help operators easily perceive system frequency response capability after different magnitudes of contingencies and at various frequency deviation levels. The FRC curve can be obtained based on data available in the control centers, including the unit on/off statuses, parameters of governors (such as deadbands, droop ratios, and headroom), and the damping of loads.

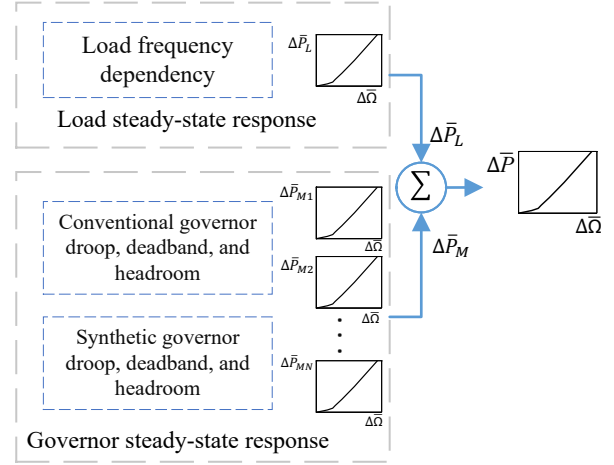

Fig. 1. System frequency response characteristic (FRC) curve generation

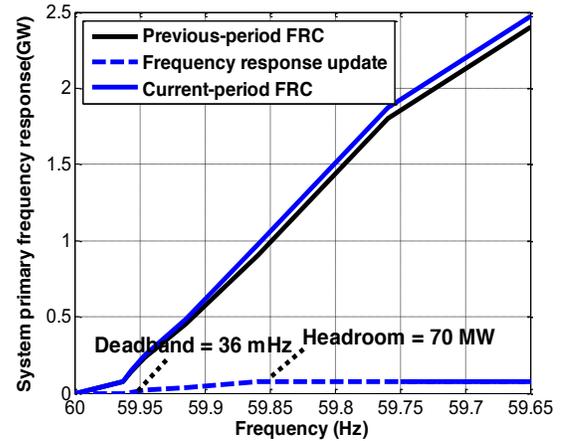

Fig. 2. System FRC curve updates

$$F_{FRC} = \sum_{i=1,\dots N} x_i \cdot f_i + f_L \qquad (1)$$

This work was supported by the U.S. Department of Energy SunShot Office SuNLaMP program under award number 30844 and also made use of Engineering Research Center Shared Facilities supported by the Engineering Research Center Program of the National Science Foundation and DOE under NSF Award Number EEC-1041877 and the CURENT Industry Partnership Program.

Shutang You is with the Department of Electrical Engineering and Computer Science, the University of Tennessee, Knoxville, TN 37996 USA (e-mail: syou3@vols.utk.edu)

where $F_{FRC}$ is the system FRC curve. $x_i$ and $f_i$ are the on/off status and each unit's FRC curve, respectively. $f_L$ is the load damping characteristic. The formulation of the system FRC curve is summarized graphically in Fig. 1. Using the operation plan of each unit, the FRC curve can also be conveniently updated in real time and predicted in short term by superimposing the frequency response characteristic of this unit onto the original FRC curve:

$$F'_{FRC} = F_{FRC} + \sum_\emptyset \Delta f_\varphi \qquad (2)$$

where $F'_{FRC}$ is the FRC curve for previous period; $\Delta f_\varphi$ is the frequency response characteristic of newly-turned-on unit $\varphi$. Taking various profiles, $\Delta f_i$ can represent any resource that provides frequency response, including synthetic governors from inverters of renewables and energy storage. Fig. 2 shows an example of updating the system FRC curve after turning on a governor-responsive unit with deadband and headroom.

### B. Fast Frequency Response Assessment (Frequency Nadir Prediction)

For power systems with obvious frequency nadir in frequency response, such as the U.S. Electric Reliability Council of Texas system (ERCOT) and Western Electricity Coordinating Council (WECC) systems, fast prediction of frequency nadir is very important for taking remedial measures to prevent under-frequency load shedding during frequency transient periods. As a transient attribute, frequency nadir prediction involves system inertia and dynamics of governors and turbines. The block diagram shown in Fig. 3 is proposed for fast assessment of frequency response and prediction of the frequency nadir. In this approach, the inertia $H$ is estimated based on the current operation plan (on/off status) submitted by each generator [10]. The aggregation of governors and turbine models is performed based on clustering governor/turbine dynamic models and associated parameters, which are largely determined by the technology type and capacities of in-service generators. For 'always-on' units (continuously operating for more than 24 hours), the clustering and aggregation are performed off line, while shoulder-load and peak-load units are modeled individually for update convenience during operation.

### III. CASE STUDIES

This case study is based on the detailed models of two interconnections of the U.S.: the Eastern Interconnection system (EI) and the ERCOT. For each system, a series of models representing high renewable scenarios have been developed [11]. The obtained FRC curves of EI in different renewable penetration scenarios are shown in Fig. 4. In this figure, the turn points of the FRC curves near 59.964 Hz (the dash line) reflect the effects of governor deadbands on system frequency response (0.036 Hz is the common deadband value in the EI system). The green circles represent the EI steady-state frequency obtained using the full dynamic simulation and applying different contingencies. These results show that FRC curve can provide operators an accurate picture of system frequency response capability adequacy at the full frequency band.

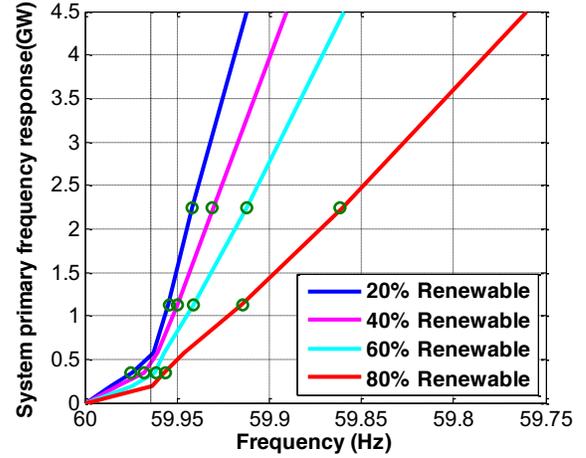

Fig. 4. The EI system FRC curves in various renewable penetration scenarios

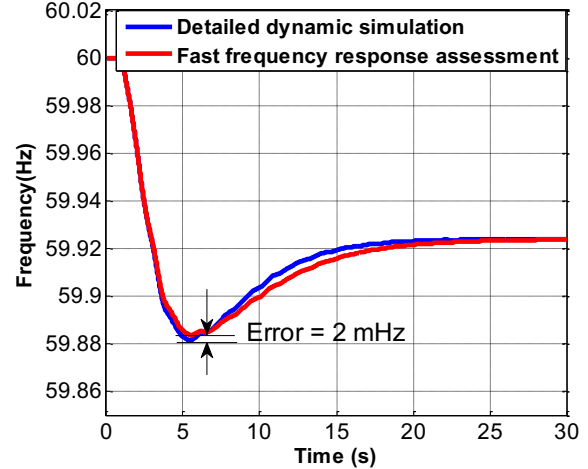

Fig. 5. Fast frequency response assessment (frequency nadir prediction) in ERCOT (200MW loss, 60% renewable scenario)

With less inertia compared with the EI, the ERCOT shows an obvious nadir in frequency response, which is a focus point of ERCOT operators. Fig. 5 is a comparison of the fast frequency response assessment result and the detailed model result of ERCOT. It shows that the proposed model can accurately predict the frequency nadir, alerting the potential need for under-frequency remedial actions. In addition, as a supplementary of the proposed FRC curve, which addresses frequency response capacity adequacy, the proposed model can help operators and planners evaluate the impact of deployment time of frequency response resources, which is critical for low-inertia systems.

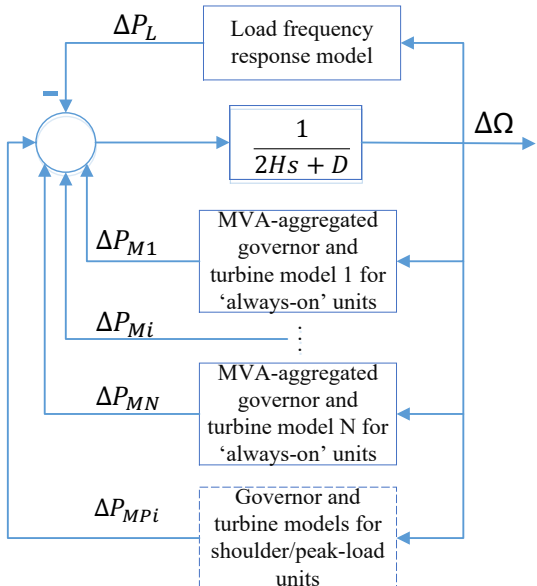

Fig. 3. Fast frequency response assessment (frequency nadir prediction) diagram

## IV. Conclusions

This letter introduces a frequency response characteristic (FRC) curve and its application in high renewable power systems. In addition, the letter presents a method for fast frequency response assessment and frequency nadir prediction without performing dynamic simulations using detailed models. The effectiveness of the proposed technology in predicting the frequency nadir is verified in the ERCOT study system. The proposed FRC curve and fast frequency response assessment method are useful for operators to understand frequency response performance of high renewable systems in real time.

## References

1. Guo, J., et al. *An ensemble solar power output forecasting model through statistical learning of historical weather dataset.* in *2016 IEEE Power and Energy Society General Meeting (PESGM)*. 2016. IEEE.
2. Hadley, S., et al., *Electric grid expansion planning with high levels of variable generation.* ORNL/TM-2015/515, Oak Ridge National Laboratory, 2015.
3. Hadley, S.W. and S. You, *Influence Analysis of Wind Power Variation on Generation and Transmission Expansion in US Eastern Interconnection.*
4. Sun, K., et al., *A Review of Clean Electricity Policies—From Countries to Utilities.* Sustainability, 2020. **12**(19): p. 7946.
5. You, S., *Electromechanical Dynamics of High Photovoltaic Power Grids.* 2017.
6. You, S., et al., *Co-optimizing generation and transmission expansion with wind power in large-scale power grids—Implementation in the US Eastern Interconnection.* Electric Power Systems Research, 2016. **133**: p. 209-218.
7. Virmani, S. *Security impacts of changes in governor response.* in *IEEE Power Engineering Society. 1999 Winter Meeting (Cat. No. 99CH36233)*. 1999. IEEE.
8. Mohajeryami, S., et al. *Modeling of deadband function of governor model and its effect on frequency Response characteristics.* in *2015 North American Power Symposium (NAPS)*. 2015. IEEE.
9. Miller, N., et al., *Eastern frequency response study*. 2013, National Renewable Energy Lab.(NREL), Golden, CO (United States).
10. Du, P. and J. Matevosyan, *Forecast System Inertia Condition and Its Impact to Integrate More Renewables.* IEEE Transactions on Smart Grid, 2017.
11. You, S., et al., *Impact of high PV penetration on the inter-area oscillations in the US eastern interconnection.* IEEE Access, 2017. **5**: p. 4361-4369.